# Friedel oscillations responsible for stacking fault of adatoms: The case of Mg(0001) and Be(0001)

*Alcántara Ortigoza, Marisol [a] [*]; Aminpour, Maral [b]; and Rahman, Talat S [c].*
*Department of Physics, University of Central Florida, Orlando, Florida 32816-2385, United States of America*

[a] Corresponding author electronic address: Marisol.AlcantaraOrtigoza@ucf.edu
[b] Electronic address: maral@knights.ucf.edu
[c] Electronic address: Talat.Rahman@ucf.edu

**Abstract**

We perform a first-principles study of Mg adatom/adislands on the Mg(0001) surface, and Be adatom on Be(0001), to obtain further insights into the previously reported energetic preference of the fcc faulty stacking of Mg monomers on Mg(0001). We first provide a viewpoint on how Friedel oscillations influence ionic relaxation on these surfaces. Our three-dimensional charge-density analysis demonstrates that Friedel oscillations have maxima which are more spatially localized than what one-dimensional average density or two-dimensional cross sectional plots could possibly inform: The well-known charge-density enhancement around the topmost surface layer of Mg(0001) is strongly localized at its fcc hollow sites. The charge accumulation at this site explains the energetically preferred stacking fault of the Mg monomer, dimer and trimer. Yet, larger islands prefer the normal hcp stacking. Surprisingly, the mechanism by which the fcc site becomes energetically more favorable is not that of enhancing the surface-adatom bonds but rather those between surface atoms. To confirm our conclusions, we analyze the stacking of Be adatom on Be(0001) – a surface also largely influenced by Friedel oscillations. We find, in fact, a much stronger effect: The charge enhancement at the fcc site is even larger and, consequently, the stacking-fault energy favoring the fcc site is quite large, 44 meV.

## I. Introduction

In bulk magnesium, as well as in those metals bound together predominately by metallic bonds, the delocalized and loosely bound valence s-electrons ($3s^{2-x}\ 3p^x$ - electrons in the case of Mg) find themselves in a nearly constant potential and thus behave as nearly free electrons. Naturally, the surface, the matter-vacuum interface, is a perturbation to that constant potential and is screened by those nearly free electrons. As a result of this screening – chiefly controlled by the electrons around the Fermi level – charge density oscillations are formed within the metal perpendicular to the surface and are damped away from the surface (i.e., the perturbation) toward center of the slab. These are long-range standing waves of period $\lambda_F/2$, where $\lambda_F$ is the Fermi wavelength. They are known as Friedel oscillations.

Friedel oscillations have been applied for decades to understand how nearly-free electrons screen point-like defects or impurities or, simply, how an otherwise homogeneous Fermi gas or Fermi liquid screens a defect. In alkaline-earth metals, whose surface is a perturbation, the Friedel oscillations arising in them have been extensively studied from the theoretical point of view in order to shed some light on the structural relaxations arising at their surfaces [1-5].

The charge density Friedel oscillations occurring inside a Mg slab have been corroborated in several computational studies [2-5]. Cho et al. [2] and later Staikov and Rahman [3] and Wachowicz and Kiejna [4] obtained the xy-integrated valence charge density as a function of the position perpendicular to the surface, the z-axis, and localized the maxima and minima of the Friedel oscillations in Mg(0001) slabs. In order to do that, the three groups calculated the difference between the xy-average valence charge density profile of bulk Mg and that of a bulk-terminated slab, which gives a one-dimensional picture of the Friedel oscillations. This result was then normalized over the charge density profile of the bulk. Later on, however, Wachowicz and Kiejna [5] presented a more detailed view of the charge redistribution in Mg(0001) by using 2D cross sectional charge-density difference contours.

Friedel oscillations cause a periodic accumulation and depletion of electronic charge between the layers. In the case of Be(0001) [6], Mg(0001) [7] and Al(111) [8], the period of the Friedel oscillations in them ($\lambda_F/2$) is such that a charge accumulation falls around the position of the first layer. The one-dimensional analysis of the Friedel oscillations for Mg(0001) [3] in particular, shows that the period of the charge density oscillation is such that its first and largest peak corresponds to a charge accumulation very close to the position of the first layer and charge depletion around the top-most interlayer separation. Moreover, although not discussed or mentioned in their work, the two-dimensional plots of Wachowicz and Kiejna hint that the first charge-density Friedel oscillation peak depicted in the 1D view [2-4] lie in the interstitial space between the atoms at the surface. Yet, as we shall see, a three dimensional (3D) inspection of the Friedel oscillations, which is still lacking, is needed to locate exactly the region of the interstitial space that is holding that "extra" charge and thus understand its effect on the binding of adatoms and their diffusion.

Friedel oscillations have been proposed as the cause of the expansion of the topmost inter-layer distance of Mg(0001) on grounds of electrostatic repulsion. Feibelman, on the other hand, has provided a chemical explanation for the relaxation of hcp(0001) surfaces based on the manner that the surface modifies the occupation of the $3s^{2-x}$ and $3p^x$ states and thus promotes the separation of the first layer [9]. In this work, we will show that the idea that the expansion is caused by Friedel oscillations is in the same context as the chemical view [9] but the current understanding on how Friedel oscillations relate to the outward expansion needs to be revised. In fact, the chemical view provides the mechanism by which the Friedel oscillations are formed.

In an earlier study [10], we have reported the relatively and surprisingly large energy (15 meV) favoring the stacking fault of Mg adatoms; i.e., the preferred adsorption at the "infinite hollow" fcc site. It is thus intriguing to relate the rationale for such preference and the large stacking-fault energy to the (so far) unclear exact localization of the first and largest peak of the Friedel charge-density oscillations. In this work, we suggest on the basis of our findings, that the preferred stacking is linked to the Friedel oscillations. To pursue this idea, we shall first find the location around the surface where the charge accumulation/depletion takes place. We shall then delineate whether the preference for the stacking fault persists for adislands and trace a rationale underlying that by analyzing the local geometric structure and charge density. We shall then corroborate our conclusions by testing them on Be(0001), which also displays a large expansion of the first topmost interlayer distance, and on a Be adatom on Be(0001), which to our knowledge has not yet been investigated. The manuscript is organized as follows: Section II summarizes our computational methodology. Section III contains our results and discussion. Finally, Section IV outlines our main conclusions.

## II. Computational details

We perform periodic density-functional-theory calculations of the total energy and the electronic structure, as implemented in the computational code QUANTUM ESPRESSO [11]. The interaction between ions and electrons is described by ultra-soft Vanderbilt pseudo-potentials with 2p-semicore states [12]. For the electron exchange-correlation functional, we have used the parameterization of Perdew and Wang 91 (PW91) based on the generalized gradient approximation (GGA) [13].

The electronic wave functions were expanded in a plane-wave basis set with a kinetic energy cut-off of 476 eV. The charge density Fourier expansion is truncated at 5440 eV. The Mg(0001) and Be(0001) surfaces were modeled by using slabs of 18 layers with a 1x1 in plane periodicity in order visualize the charge density Friedel oscillations of Mg(0001). For modeling Mg adislands on Mg(0001), we have used 7-layer films with an in-plane periodicity of 3x3 atoms. Integrations inside the Brillouin zone were performed by sampling the latter according to the Monkhorst–Pack scheme [14] with a uniform grid of 16x16x1 and 5x5x1 k-points for the 1x1 surface and the 3x3 surface, respectively. Integrations use the Gaussian broadening technique for the level occupation with a smearing parameter of 0.1 eV. For convergence test, we refer the reader to Ref.[10].

In all calculations involving a surface, a vacuum space of 20 Å separates the periodic images of the slab to avoid interaction between them. The positions of all atoms in the slab were optimized until the Hellmann–Feynman forces on each atom and each direction was smaller than $2.57 \times 10^{-3}$ eV/Å.

Charge densities differences are evaluated as $\delta\rho = \rho[Mg_n/Mg(0001)] - \rho[Mg(0001)] - \rho[Mg_n]$, where $\rho[Mg_n/Mg(0001)]$ is the charge density of the entire system (n-mer plus the surface) with ion-cores in their relaxed configuration, $\rho[Mg(0001)]$ is the charge density of the clean surface and $\rho[Mg_n]$ is the charge of the isolated n-mer, $Mg_n$. Note that in order to obtain a consistent charge densities difference, the positions of the atoms used to compute $\rho[Mg(0001)]$ and $\rho[Mg_n]$ are extracted from those in the fully relaxed *composite* systems and not from the actual relaxed coordinates of clean Mg(0001) and free-standing Mg n-mer.

## III. Results and Discussion

We have verified elsewhere [10] that the applied methodology is accurate enough to reproduce properties of bulk Mg and Mg(0001). The lattice parameters (a= 3.213 Å and c/a =1.607 Å), cohesive energy (1.45 eV) and bulk modulus (35.48 GPa) of bulk Mg are in very good agreement with the experimental values (a=3.21 Å and c/a=1.624, $E_{coh}$=1.46 eV) [15] and with the previous calculations [3-5,16,17]. The surface energy as well as the contraction/expansion of the interlayer distance among the topmost layers is also found to be in good agreement with experimental values. Further details can be found in Ref.[10].

### A. New insights into the charge density Friedel oscillations

*1. The outward relaxation of Mg(0001):* Before turning to new insights into Friedel oscillations available through a three-dimensional analysis of the charge density, some remarks about the current understanding of the relationship between Friedel oscillations and interlayer expansion are in order. Cho et al. [2] proposed that relaxations of ions at metal surfaces arise as a response to Friedel oscillations, which cause accumulation and depletion of charges within the crystal. In this scenario, atoms of an entire layer may be left effectively charged, giving rise to an electrostatic interaction among layers. Thus, the relaxation of the surface may depend on whether the electrostatic force between layers is attractive or repulsive [3]. This argument has been applied to explain the relaxation of Mg(10$\underline{1}$0), Be(10$\underline{1}$0) [2], Mg(0001) [3] and Be(0001) [4,5]. Based on one-dimensional charge-density profiles, it has been argued that atoms constituting the first, second and third layers are effectively negatively charged. Hence, according to this interpretation of the charge density difference, these ionic layers repel, causing the well-known interlayer expansion. However, some scrutiny of this interpretation is called for: In systems composed of atoms of the same species ionic charge transfer is not possible as the atoms have the same electronic affinity and thus ionicity cannot be a distinguishing feature describing their bonding. Moreover, were the atoms effectively negatively charged, we should find positively charged atoms somewhere else for the system to be neutral, just as in any ionic bonding. Say, if the first four Mg layers were effectively negatively charged, one would need other Mg layers to be positively charged (by the same magnitude). In reality, one-dimensional profiles of Friedel oscillations in Mg(0001) [2-4] do not imply that the atoms are effectively charged; they indicate that, since the first and largest peaks of the charge-density Friedel oscillations in Mg(0001) coincide with the position of the first atomic layers, upon surface formation the bonding charge *abandons* the interlayer space and becomes more localized around the atoms forming the first layers. Thus, from the available one-dimensional average density, one could only construe that the layer might be approaching a free-standing monolayer condition. In summary, the layers are not effectively charged: The effect of the Friedel oscillations is to reduce the interlayer bonding charge. Naturally, this weakens the (metallic) interlayer bonds and causes the first three layers to separate from each other. The latter is further evidenced by a Bader analysis of the Mg slab. It shows that all atoms below the surface layer have the same charge. The first-layer atoms in fact appear slightly positively (rather than negatively) charged within the accuracy of the calculation. Specifically, a small fraction of the charge of the surface atoms slowly decays and spreads into the vacuum but not all the charge in the tail can be taken into account. Finally, this interpretation of how the Friedel oscillations cause the outward relaxation of Mg(0001) does not rule out the chemical view [9]. Specifically, Friedel oscillations reduce interlayer charge and increase intralayer charge (Figs.1(c) and 1(d)), but the interlayer bonding charge may correspond to depletion of $p_z$ states whereas intralayer charge may correspond to the population of *s*-states, which makes surface atoms more free-atom-like, i.e., with a charge distribution more spherically symmetry, as implied by Figs.1(c) and 1(d)).

Let us now turn to the exact location of the charge accumulated around the first layer. To this end, we plot in Figure 1(a) the difference between the charge of bulk Mg and that of a 18-layer slab – as in the previous calculation – but this time in three dimensions. First of all, in agreement with Ref. 3 and 4, we find that the displaced charge in Mg(0001) slabs is indeed located mainly around the position of the first layer. Then, as shown in Ref. 4, the charge is localized around the interstitial space between the atoms at the surface. Yet the present 3D difference isosurfaces reveal in addition that the displaced charge *lies at the fcc infinite-hollow site* of Mg(0001) (red pocket in Figure 1(a)).

Fig.1(b) in turn shows the [0001] cross section of the charge-density-difference isosurfaces in part (a). The latter two-dimensional view allows us to see that the charge **excess** extends up to the bridge, whereas the charge around the hcp site is slightly reduced. We shall see that the latter features have implications on the Mg adatom binding energy and its diffusion.

A word of caution has to be given before moving to considerations of the binding of adatoms. Note that, in order to capture Friedel oscillations in metallic slabs, previous and present calculations [Figs.1(a) and (b)] address differences between the charge density of bulk Mg and that of a *bulk-terminated* slab. However, Friedel oscillations are long-ranged and their wavelength is not perfectly commensurable with the interlayer distance (a feature reflected in the slower convergence of relatively deep subsurface interlayer distance [10]). Therefore, the charge throughout the film can vary once the forces on the atoms of a bulk-terminated slab are relaxed and nothing guarantees that the charge density enhancement remains as depicted in Figs. 1(a) and (b). This uncertainty is of particular concern if we want to understand the effect of the Friedel oscillations on adatom binding and diffusion barriers. Therefore, we contrast the [0001] cross section at the surface layer of the total charge density of a *totally relaxed Mg slab* (Fig.1(c)) and that of the bulk layer (Fig.1(d)). These two plots demonstrate that the charge density around the surface (totally relaxed) is indeed significantly enhanced around the fcc site with respect to the charge density of bulk layers even after full force relaxation.

**B. Influence of Friedel oscillation on adsorption of Mg adatom and adislands on Mg(0001)**

In this section, we shall see that the charge accumulation around the fcc site, which is caused by Friedel oscillations is of consequence for the binding and stacking of small Mg adatom islands on Mg(0001). Before doing so, we provide the energetics of the adsorption of the adislands and an analysis of their structure. Next, we scrutinize the charge density distribution to locate the features responsible for such preference. In fact, that the enhanced charge density around the first layer -- derived from the Friedel oscillations -- lies precisely at the fcc site suggests that the "extra" charge density pocket (Fig.1(a)) contributes to charge density bonding the adatom. Still, such explanation begs the question: why would those factors promoting the stacking fault of the monomer stop operating as the adisland reaches the size of a tetramer or as the adislands approach each other? We shall see that although the "extra" charge density pocket does cause the stacking fault preference, its role is not simply that of strengthening the bonds of the adatom.

**1. Energetics**

The preference for the fcc stacking fault is not exclusive to the monomer. We have calculated the binding energy per atom ($E_B$) of several Mg adislands – from monomer to octamer – and of a full overlayer placed at both the fcc and the hcp sites. For the dimer, trimer, tetramer and hexamer several configurations and orientations were tested, while for larger islands computational costs refrained us from testing more than one configuration. Table 1 summarizes our calculated energetics of adislands at the fcc and hcp sites (the data for the most stable dimer and trimer). It shows that the preference for the fcc stacking fault persists at least up to the trimer. Furthermore, the behavior of these small adislands is not qualitatively dependent on whether the position of substrate atoms is relaxed or rigid.

We cannot rule out that larger islands also display a preference for stacking fault because in our calculation, as the adislands grow (tetramer, hexamer, heptamer and octamer), they necessarily interact with each other and favor again the hcp site (Table 1). In fact, a full layer prefers the hcp site over the fcc stacking fault by 15 meV per atom (See Table 1). In fact, our calculations suggest that as the islands get closer to each other the preferred site is again hcp. In our particular supercell setup, the turning point between fcc and hcp preferred binding corresponds to a coverage between one-third and one-half monolayer.

## 2. Structural characterization of Mg$_n$ adislands on Mg(0001)

We now turn to investigate the origin of the preference of the fcc stacking fault. In pursuing this aim, we shall examine the adislands that display this preference (from monomer to trimer) and the one at the turning-point for the hcp preference, the tetramer. As a preliminary step, we analyze the bond-length of Mg adatom/adislands adsorbed on Mg(0001) under two conditions: when the whole system is allowed to relax and when the substrate is kept frozen. The structural characterization is presented in Tables 2 and 3, respectively. However, upon analyzing the data, we find that neither the distances among the adisland atoms, nor the height of the adisland with respect to its substrate nearest neighbors (NN), nor the distances among the adisland NNs provide a hint about the mechanism behind the stacking fault preference or establish a consistent bond-order trend. For example, the data for the monomer could in principle indicate a slight tendency of the adatom to stay farther from its substrate neighbors (weaker bond) when sitting at the hcp site than when sitting at the fcc site. However, the opposite trend is displayed by the trimer whether the substrate is relaxed or not. The structure of the dimer on Mg(0001) does not provide much insight either because the stacking fault preference makes the hcp sites unstable. In the totally relaxed system, the dimer spontaneously slides toward the bridge, almost reaching fcc sites. As a result, it displays two relatively short bonds and two bonds that are significantly longer than the lattice parameter, a. Allowing both the dimer and the substrate to relax from the hcp sites toward the bridge-like configuration reduces the fcc stacking fault preference to 4 meV (Table 1). However, the latter stacking-fault energy is with respect to the bridge-like configuration, which is a local minimum. Now, if the substrate is frozen, the hcp configuration of the dimer is stabilized but the hcp site becomes less favorable than the fcc site by an even larger energy per atom. In the case of the tetramer, the related bonds are so spread out that it is not possibly to draw any conclusion. Clearly, no argument for site preference of the islands can be built upon considerations of bond lengths.

## 3. Charge density analyses

Attaining an understanding of the charge density distribution responsible for the preference of small islands to sit at fcc rather than hcp sites is nontrivial. Actually, it would not be reasonable to trace the answer via charge density differences (as those shown for the clean surface in Fig.1 (a) and (b)), since these difference are of the order of few meV and probably smaller than those caused by inherent errors in the charge density differences analysis that follows from our discussion in Sec. II about the ionic configurations under which the total energies of the systems are calculated. In fact, such analysis does not provide rational for the stacking fault. The only option is thus to investigate the total charge density of the composite system, $\rho[Mg_n/Mg(0001)]$. This is a reasonable approach since the Hohenberg-Kohn theorem tells us that if small Mg adislands on Mg(0001) prefer the fcc site over the hcp site, this preference must necessarily be reflected in the charge density distribution. In order to dispel the idea that local relaxations largely influence our analysis, we include the case when the $Mg_n/Mg(0001)$ system is fully relaxed and that when the substrate is kept frozen in our analysis. Importantly, as shown in Figure 2, both the totally relaxed and the bulk-terminated Mg(0001) surface display qualitatively the same landscape to the adatom, providing us a simpler scenario for our analysis.

The next feature of the total charge density that calls attention is that fcc sites render less charge density than the hcp sites, despite the "extra" charge density pocket located at the fcc site, as shown by our total charge density plots (Fig. 2). This implies that the reason for which the adatoms prefer the fcc site is not an increased availability of charge to make the bond.

Visualizing the total valence charge involved in the bonds is also not free of challenges. On the one hand, three-dimensional plots do not reveal a charge density profile within a charge interval but isosurfaces for a fixed charge density value. On the other hand, turning to analyze two-dimensional cross-sectional charge density profiles (e.g. planes parallel to the surface) is not straightforward because comparisons ought to be made between the monomer at fcc and that at hcp, then between the monomer cases and the dimer cases and so on, but in each of these cases the height of the adisland (A) with respect of the surface atoms (S), $Z_{AS}$, varies significantly, as shown in Table 2, and a fair comparison could be compromised. Nevertheless, since both the totally relaxed and the bulk-terminated Mg(0001) display qualitatively the same landscape to the adatom and the stacking fault energy trend is also qualitatively the same, we can grasp the essentials of the charge density distribution responsible for the stacking fault preference in small adislands by considering the n-mers first on the frozen bulk-terminated Mg(0001). We have thus relaxed only the n-mers (n=1,..,4) at the fcc and hcp sites on a bulk-terminated Mg(0001). In this case, the adislands heights do not vary as much as for the totally relaxed system (See $Z_{AS}$ in Table 2 and 3), which allows us to make a meaningful comparison: We compare two-dimensional charge density profiles of all the adislands for planes at exactly the same height with respect to the substrate. The two-dimensional charge density profiles of $\rho[Mg_n/Mg(0001)]$ described above (in which Mg(0001) is bulk terminated) in Table 4 reveal that when the monomer sits on the fcc site, the bonds among its NN substrate atoms are strengthened, rather than those between the monomer and its substrate NN atoms. Table 4 also demonstrates that while the same effect is displayed by the fcc dimer, it not so for the hcp monomer or the hcp dimer. In the case of the trimer, the charge density profiles at hcp and fcc are more complex and less distinct. However, by analyzing each of the bonds to neighboring atoms in the substrate, one finds that the latter are furnished with more charge density when the trimer sits at fcc sites than when it sits at hcp sites. For the tetramer, when the adislands start to interact strongly, only subtle features might indicate a more energetically favorable configuration at the fcc site. Not surprisingly, the stacking fault energy becomes very small (Table 4). The height of the charge density plane (at 1.2 Å above the surface) was taken to lie between the adatom/adisland and the surface atoms in the above analysis. The position of the plane was easily chosen because the charge profiles for a given n-mer are very similar at fcc and hcp site except around the position of the plane shown in Table 4.

Once we have identified the feature of the charge density redistribution that could account for the preference of the fcc stacking fault, we can proceed to trace the same features in the two-dimensional profiles of $\rho[Mg_n/Mg(0001)]$ when the entire system is allowed to relax. To our surprise, similar features appear for the totally relaxed system and at practically the same distance from the substrate atoms (~1.2 Å, See Table 5). Specifically, the monomer at fcc site also induces a charge density enrichment in the bonds between its NNs and other neighboring atoms that does not appears when the monomer sits at the hcp site. In the case of the dimer, one sees that the dimer at the fcc site also induces a charge density enrichment in the bonds between its NNs and other neighboring atoms, yet, the same happens when the dimer is at the bridge (rather than hcp) position and at a larger extent. So, the fact that it is unstable at the hcp site ruins any possible comparison. Nevertheless, the features in the charge density when the dimer sits at the hcp, although not adding to the supporting evidence, are not enough to rule out our argument. Namely, the strong enrichment of the bonds between the dimer's NNs and other neighboring atoms occurs also at the expense of, or accompanied by, broken bonds between the dimer and two of its two surface neighbors (See Table 5). Overall, the energy associated to the fcc stacking fault preference is reduced significantly to 4 meV per atom. The trimer at fcc sites, in contrast, is favored by as much as 10 meV/atom with respect to the hcp sites. The trimer at fcc sites also induces a charge density enrichment in the bonds between its NNs and other neighboring atoms that is larger than that occurring when the trimer sits at the hcp sites. Finally, the tetramer at fcc sites is favored by only 1 meV/atom with respect to the hcp sites. As we turn to the tetramer, one sees that the bonds between its NNs and other neighboring atoms are furnished by extra charge density at fcc and hcp sites almost evenly. In fact, at the hcp site, the charge enrichment on the substrate bonds is slightly larger, which coincides with the fact that the tetramer is the turning point for the preference of the stacking fault.

Based on the above analysis, we propose that the role of the charge-density pocket and is that of strengthening the bonds among substrate atoms. That is the reason for which the fcc site is preferred over the hcp. In other words, the extra charge density pocket at the fcc site tends to be distributed among the surface atoms enhancing their mutual binding. A similar behavior happens for the hcp adislands but only when they approach each other. The images in Tables 4 and 5 also provide a rationale for the decline in the preference for the fcc stacking fault as the adisland grow larger and/or coalesce. For example, by comparing the charge density profile between the clean surface (Fig.2(b)) and that of the monomer at fcc (in Table 5), one sees that upon adsorption of the monomer, the charge density at neighboring fcc sites is reduced, indicating that the enhanced bonding among substrate atoms does not withdraw charge exclusively from the site where the monomer sits but also from neighboring fcc sites. We find the same trend by comparing neighboring fcc sites of the monomer environment with those of the dimer: Again, the neighboring fcc sites of the dimer become more depleted. Similarly, for the trimer and tetramer at fcc, one clearly sees that neighboring fcc sites get even more charge-depleted. Finally, notice that the redistribution and spreading of the charge upon the adsorption of adatoms (the charge brought at the fcc sites by the Friedel oscillation) should in fact contribute to reduce the electronic kinetic energy.

**C. Friedel oscillations in Be(0001) and their effect on adatom adsorption**

Up to now we have shown the features of the charge density distribution responsible for the preference of small islands to sit at fcc sites rather than at hcp sites. However, this is only indirect evidence that the stacking fault is caused by the Friedel oscillations. We thus find necessary and opportune to strengthen our argument by testing another material. We shall thus consider Be(0001), another hcp sp- and nearly-free-electron metal that also display Friedel oscillations and whose relaxation of the topmost interlayer distance yields an expansion (See Ref.4). In this section we are concerned about three issues of Be(0001): (1) whether the topmost maximum of the Friedel oscillations in Be(0001) is also localized around the first layer; (2) whether Friedel oscillations also induce a charge accumulation precisely at the fcc site, and (3) whether the Be monomer also prefers to bind at the faulty fcc site of Be(0001). The answer to the first question is provided in Ref.[4]: Be(0001) also displays the maxima of the Friedel oscillations at the first layer. Notice that the maxima of the Friedel oscillations are apparently less conspicuous than those of Mg(0001) (See Fig.1 of Ref.4). The reason is that they divide the charge density differences by the charge density of the bulk in order to present a normalized value. As a result, Fig.1(b) of Ref.4 does not anticipate that the Friedel oscillation maxima are, in absolute value, larger than those of Mg, since Be holds bonds ~30% smaller than those of Mg and, therefore, the valence charge density binding Be atoms is in general much larger than that in Mg. In Figure 3, we compare the Friedel oscillations in Mg(0001) and Be(0001) via a three dimensional charge-density difference. Clearly, the charge density enhancement for Be(0001) is dramatically larger than that found for Mg(0001). More importantly, the charge enhancement is localized at the fcc site, as in Mg(0001), which answers our second question. Even more importantly, the Be monomer on Be(0001) indeed prefers the fcc stacking fault site than the hcp site by a strikingly large stacking-fault energy of 44 meV. The magnitude of this value may further support the argument that the stacking fault preference is driven by the charge-density pocket because it correlates with the magnitude of the charge enhancements. To our knowledge this is the first time that the stacking fault of Be adatoms on Be(0001) is reported, as well as the mechanism responsible for it. Nonetheless, the fact that the preference of adatoms for the stacking fault is driven by charge redistribution/enrichment among substrate atoms is in accord with previous studies pointing out that surface states mediate inter-adatom interactions [18].

**D. Influence of Friedel oscillations in Mg(0001) on adatom diffusion**

First principles calculations [10], in contrast to results from effective medium theory [19], indicate that the adatom diffusion barrier is quite asymmetric. That the barriers are equal according to effective-medium theory calculations implies that neither of the two sites (fcc or hcp) is energetically preferred. This is understandable because such calculations cannot model Friedel oscillations. Thus, Friedel oscillations are also responsible for the asymmetry of the diffusion barrier [10]. Namely, they cause a charge accumulation at the fcc site and a charge depletion at the hcp site, which stabilizes the fcc adsorption site and destabilizes the hcp site (Fig. 1(b)). The fact that the barriers are so low is of course related to the fact that Mg bonds are rather weak compared to most metals. We shall see, however, that the Friedel oscillations are also related to the height of the barrier. Not only do they modulate the binding energy of the two local minima (hcp and fcc), as described above, but also tune the energetics along the diffusion path. Namely, although the extra charge-density pocket caused by the Friedel oscillations is strongly localized at the fcc site, the charge enhancement at the surface extends well up to the bridge site; Fig.1(b) and comparison of Figs.1(c) and (d) clearly show that. Since furnishing charge at the bridge site smoothens the potential energy surface for the adatom by increasing the binding energy around the transition state, Friedel oscillations also influence the energy barrier. In summary, Friedel oscillations promote Mg adatom diffusion with very short transit time at hcp sites.

**IV. Conclusions**

We report a first-principles study of Friedel oscillations in Mg(0001) and Be(0001) and their effect on the binding and diffusion of adatoms. We provide a new viewpoint on how Friedel oscillations influence the relaxation of these surfaces that is compatible with the changes found in the occupation of its *s* and *p* states. Our three-dimensional charge-density analysis reveals key details of the maxima/minima of the Friedel oscillations depicted previously by one-dimensional charge-density difference plots: We find that the long-known charge-density enhancement at the surface of Mg(0001) and Be(0001) resulting from Friedel oscillations is strongly localized at the fcc hollow site. In turn, our analysis of the total charge density shows that the charge accumulation site favors the adsorption of the Mg monomer, dimer and trimer at the fcc "infinite" hollow, which constitutes a faulty stacking of these small islands. Yet, larger Mg islands and/or higher coverage on Mg(0001) prefer the normal hcp stacking. We show that the preference for the fcc stacking fault of small adisland is explained by the following mechanism: the enhanced charge-density at the fcc site does not strengthens the adatom-surface bonds but rather the inter-substrate bonds. In other words, this charge gets distributed among the surface atoms enhancing their mutual binding. Our findings for Mg/Mg(0001) are confirmed by Be/Be(0001): We find that the charge-density enhancement at the surface of Be(0001) is also localized at the fcc site and that is even larger in magnitude than that on Mg(0001). We report a strikingly large stacking-fault energy for Be monomer on Be(0001) of 44 meV, a value that is consistent with a larger charge accumulation at the fcc site of Be(0001). Finally, our calculations also suggest that the Friedel-oscillations-driven charge accumulation on Mg may also be responsible for the relatively small diffusion barriers for the monomer.

**Acknowledgements:**

The work is supported by US Department of Energy under Grant No. DE-FG02-07ER46354. M. Alcántara Ortigoza is indebted to Sergey Stolbov for countless insightful discussions. The authors are grateful to Michael Tringides, Zdenka Chromcová and Zdeněk Chvoj for directing our attention to the problem. We acknowledge the University of Central Florida Stokes Advanced Research Computing Center and the Teragrid partnership for providing computational resources and support.

**Table captions**

Table 1: Binding energy, $E_B$, and stacking fault, $\Delta E_B$, per atom of Mg adislands on Mg(0001) – from a monomer to an octamer – and of a full overlayer placed at both the fcc and the hcp sites for a structure in which (a) the whole system is totally relaxed (b) only the Mg adisland atoms are allowed to relax but the Mg(0001) substrate is kept frozen.

Table 2: Structural characterization of the Mg adislands -- dimer, trimer and tetramer – on the Mg(0001) surface. These values correspond to the case in which all atoms are allowed to relax. The distance among atoms forming the adislands are denoted by $d_{IA}$; $Z_{AS}$ stands for the height (vertical distance) of the atoms forming the adisland with respect to their non-equivalent substrate neighbors, $d_{NN-S}$ stands for distance between the atoms forming the island and their substrate non-equivalent nearest neighbors.

Table 3: Structural characterization of the Mg adislands -- dimer, trimer and tetramer – on the Mg(0001) surface. These values correspond to the case in which only the atoms of the adisland are allowed to relax while the atoms of the substrate are kept frozen. The distance among atoms forming the adislands are denoted by $d_{IA}$; $Z_{AS}$ stands for the height (vertical distance) of the atoms forming the adisland with respect to their non-equivalent substrate neighbors, $d_{NN-S}$ stands for distance between the atoms forming the island and their non-equivalent substrate nearest neighbors.

Table 4: (Color online) Two-dimensional plots of the total charge density of the n-mers (n=1,…,4) at the fcc and hcp sites of a bulk-like Mg(0001) substrate. The plots correspond to a plane parallel to the substrate at ~1.2 Å above it. The scale is such that dark regions denote less charge. The left-most column displays the stacking-fault energy per atom, $\Delta E_B$.

Table 5: (Color online) Two-dimensional plots of the total charge density of the totally relaxed n-mers (n=1,…,4) at the fcc and hcp sites of Mg(0001). The plots correspond to a plane parallel to the substrate at ~1.2 Å above it. The scale is such that dark regions denote less charge. The left-most column displays the stacking-fault energy per atom, $\Delta E_B$. (*) Note that, strictly speaking, the dimer does not sit at hcp sites but rather at the bridge.

**TABLES**

Table 1

| n | Totally relaxed | | | Frozen substrate | | |
|---|---|---|---|---|---|---|
| | $E_B(fcc)$ (eV) | $E_B(hcp)$ (eV) | $\Delta E_B$ (meV) | $E_B(fcc)$ (eV) | $E_B(hcp)$ (eV) | $\Delta E_B$ (meV) |
| Monomer | -0.60 | -0.58 | -15 | -0.58 | -0.57 | -14 |
| Dimer | -0.75 | -0.74 | -4 | -0.73 | -0.75 | -12 |
| Trimer | -0.75 | -0.74 | -10 | -0.74 | -0.73 | -10 |
| Tetramer | -0.78 | -0.78 | -1 | -0.77 | -0.77 | -2 |
| Hexamer | -0.92 | -0.92 | 4 | - | - | - |
| Heptamer | -0.93 | -0.95 | 11 | - | - | - |
| Octamer | -1.02 | -1.03 | 12 | - | - | - |
| Full | -1.09 | -1.11 | 15 | - | - | - |

Table 2

| adisland | $d_{IA}$ | $Z_{AS}$ | $d_{NN-S}$ |
|---|---|---|---|
| Monomer-fcc | | 2.47 | 3.12 |
| Monomer-hcp | | 2.48 | 3.12 |
| Dimer-fcc | 2.97 | 2.46, 2.64, 2.34 | 3.08, 3.09, 3.24 |
| Dimer-hcp | 2.96 | 2.46, 2.65, 2.36 | 3.04, 3.06, 3.42, 3.72 |
| Trimer-fcc | 3.06 | 2.43, 2.51, 2.51 | 3.10, 3.10, 3.13 |
| Trimer-hcp | 3.06 | 2.50, 2.50, 2.40 | 3.10, 3.10, 3.10 |
| Tetramer-fcc | 3.07, 3.08, 3.09 | 2.54, 2.42, 2.62 | 3.08, 3.14, 3.20 |
| | 3.10, 3.11, 3.11 | 2.60, 2.38, 2.38 | 3.10, 3.11, 3.11 |
| Tetramer-hcp | 3.09, 3.09, 3.13 | 2.44, 2.55, 2.36 | 3.09, 3.13, 3.19 |
| | 3.09, 3.09 | 2.60, 2.36, 2.36 | 3.08, 3.10, 3.10 |
| | 3.07, 3.07 | 2.40, 2.47, 2.47 | 3.06, 3.10, 3.10 |

Table 3

| adisland | $d_{IA}$ | $Z_{AS}$ | $d_{NN-S}$ |
|---|---|---|---|
| Monomer-fcc | | 2.53 | 3.13 |
| Monomer-hcp | | 2.53 | 3.14 |
| Dimer-fcc | 3.01 | 2.53,2.53,2.53 | 3.08,3.18,3.16 |
| Dimer-hcp | 3.02 | 2.52,2.52,2.52 | 3.06,3.16,3.17 |
| Trimer-fcc | 3.06 | 2.50,2.50,2.50 | 3.09,3.09,3.16 |
| Trimer-hcp | 3.07 | 2.49,2.49,2.49 | 3.08,3.08,3.15 |
| Tetramer-fcc | 3.08,3.08 | 2.49,2.49,2.49 | 3.13,3.14,3.20 |
| | 3.14, 3.14 | 2.49, 2.49, 2.49 | 3.05, 3.14, 3.14 |
| | 3.08, 3.08 | 2.50, 2.50, 2.50 | 3.10, 3.10,3.15 |
| Tetramer-hcp | 3.09,3.09,3.13 | 2.53,2.53,2.53 | 3.09,3.11,3.12 |
| | 3.13,3.13 | 2.48, 2.48, 2.48 | 3.03, 3.13, 3.13 |
| | 3.09, 3.09 | 2.48, 2.48,2.48 | 3.08, 3.08, 3.13 |

Table 4

| | fcc | hcp | $\Delta E_B$ (meV) |
|---|---|---|---|
| Mg Monomer | 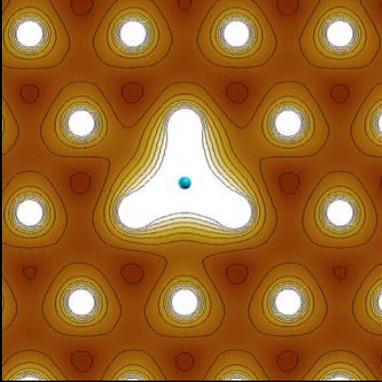 | 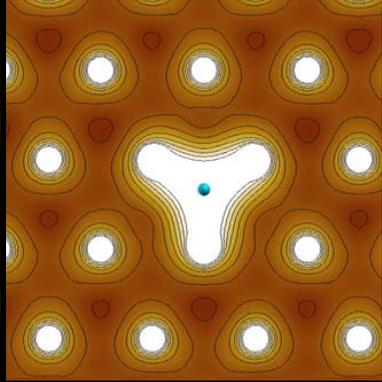 | -14 |
| Dimer | 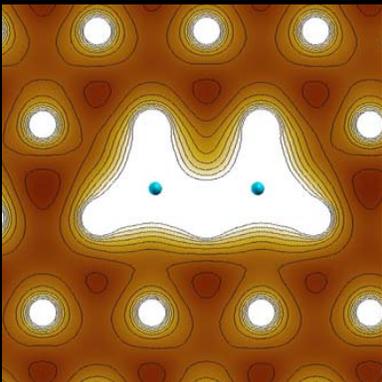 | 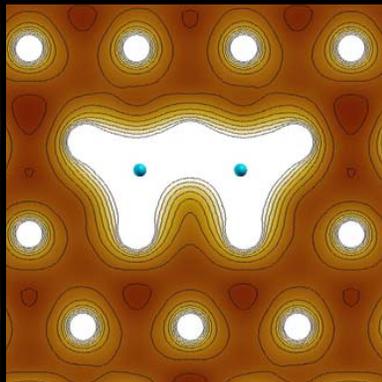 | -10 |
| Trimer | 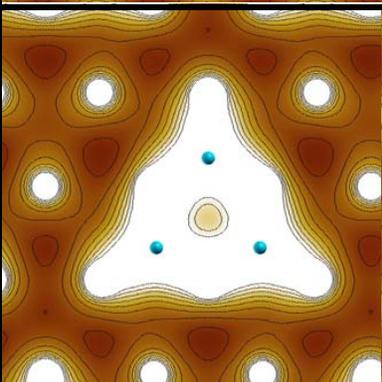 | 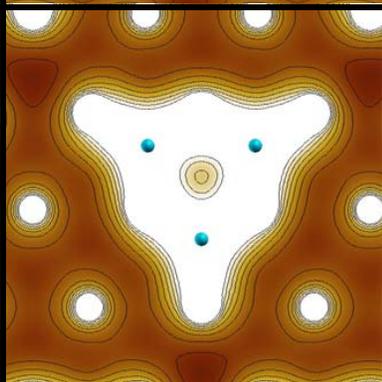 | -12 |
| Tetramer | 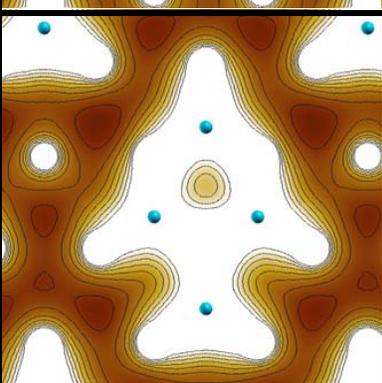 | 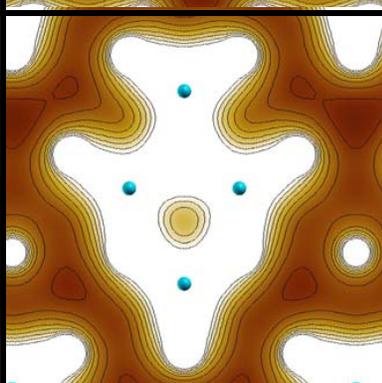 | -2 |

Table 5

| | fcc | hcp | $\Delta E_B$ (meV) |
|---|---|---|---|
| Monomer | | | -15 |
| Dimer | | | -4* |
| Trimer | | | -10 |
| Tetramer | | | -1 |

**Figure captions**

Figure 1 (Color online): (Upper insets) Difference between the charge density of bulk Mg and that of a non-relaxed bulk-terminated Mg(0001) surface (a) Isosurfaces. The z-axis is perpendicular to the surface. The (blue) balls represent the first four atomic layers of the slab. The pocket (red) indicates the region of Mg(0001) that displays more charge density than the corresponding one in bulk Mg. (b) [0001] Cross section of the isosurface in (a); the fcc region (red) displays charge accumulation and the hcp region (blue) displays charge depletion. (Lower insets) [0001] Cross sectional planes of the total charge density around (c) the fully relaxed Mg(0001) and (d) bulk Mg. Darker (brown) regions in (c) and (d) indicate less charge. In (b)-(d), the plane is located at the height of the surface or bulk atoms under consideration in order to highlight the charge accumulation around the fcc hollow site of Mg(0001).

Figure 2 (Color online): [0001] Cross section of the total charge density of (a) non-relaxed bulk-terminated Mg(0001) and (b) fully relaxed Mg(0001). Darker (brown) regions indicate less charge. The plane is located at ~1.2 Å above the position of the surface atoms.

Figure 3 (Color online): Three dimensional charge-density difference isosurfaces showing the Friedel oscillations in (a) Mg(0001) and (b) Be(0001). The charge density isovalue is the same for both surfaces. The difference is taken between the charge density of bulk Mg and that of a non-relaxed bulk-terminated Mg(0001) surface. The z-axis is perpendicular to the surface. The balls (light blue and green) represent the first three layers of the slab. The pockets (red surfaces) indicate the regions in the surface displaying more charge density than the corresponding one in bulk.

**FIGURES**

Figure 1

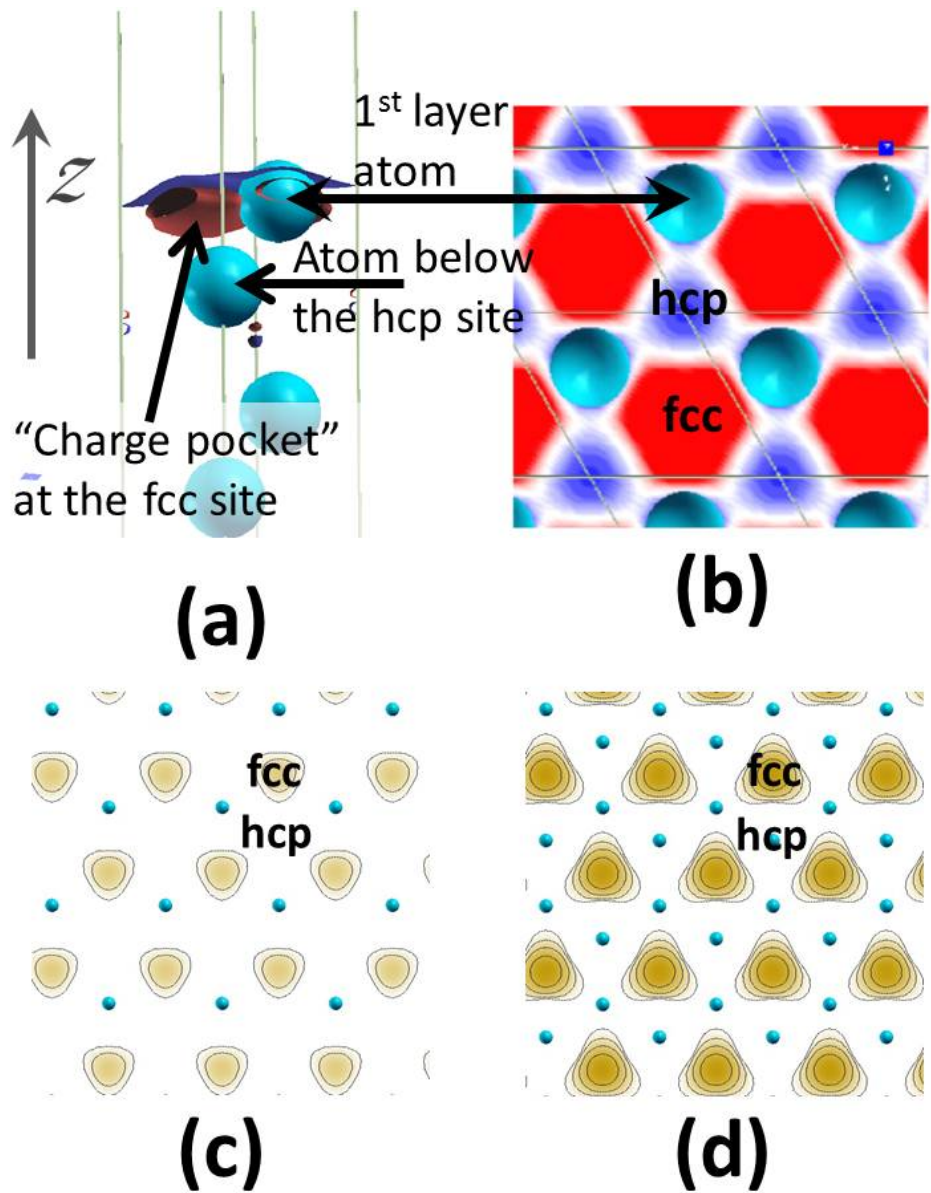

Figure 2

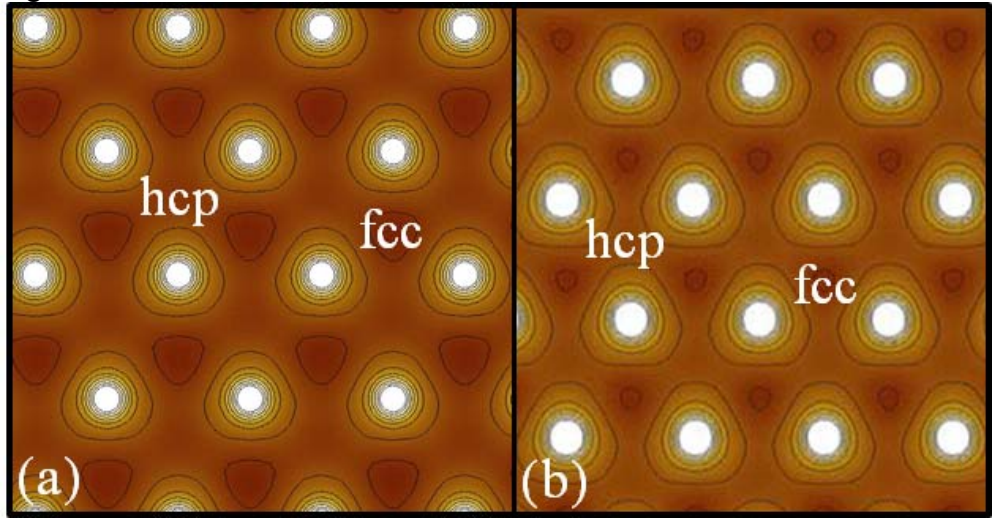

Figure 3

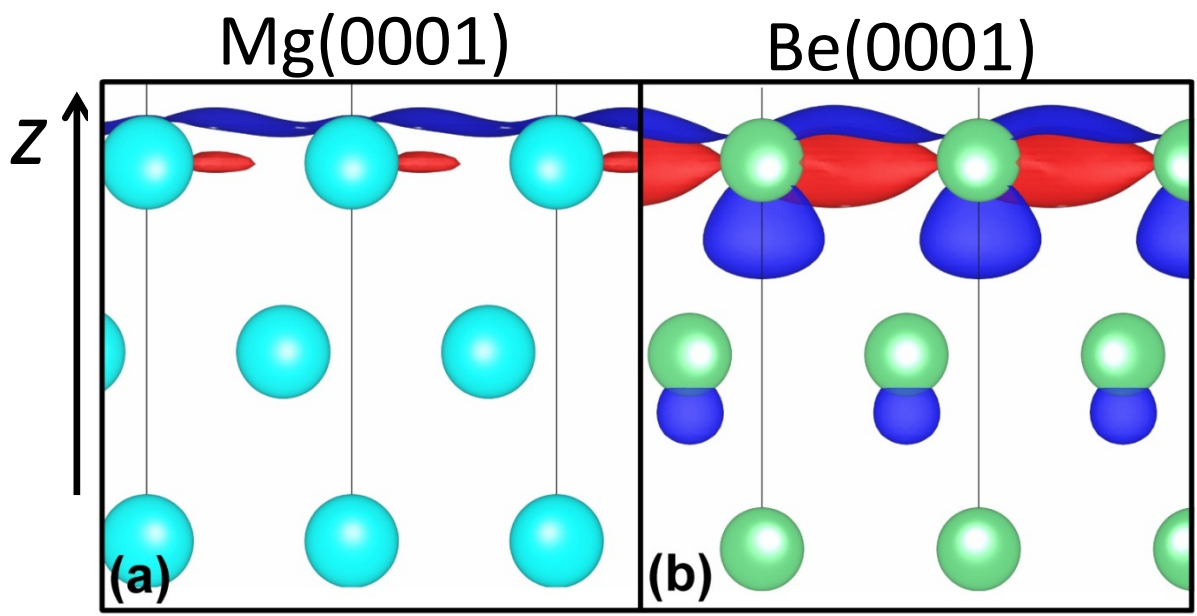